\documentclass[twocolumn]{aastex631}

\begin{document}

\received{2024 April 26}
\revised{2024 October 25}
\accepted{2024 November 11}
\submitjournal{ApJ}

\title{Radio Follow-up Observations of SN 2023ixf by Japanese and Korean VLBIs}

\correspondingauthor{Yuhei Iwata}
\email{yuhei.iwata@nao.ac.jp}
\author[0000-0002-9255-4742]{Yuhei Iwata}
\affiliation{Division of Science, National Astronomical Observatory of Japan, 2-21-1 Osawa, Mitaka, Tokyo 181-8588, Japan}
\affiliation{Mizusawa VLBI Observatory, National Astronomical Observatory of Japan,
2-12 Hoshigaoka, Mizusawa, Oshu, Iwate 023-0861, Japan}

\author{Masanori Akimoto}
\affiliation{Graduate School of Science and Technology for Innovation, Yamaguchi University, 1677-1 Yoshida, Yamaguchi-city, Yamaguchi 753-8512, Japan}

\author[0000-0002-6916-3559]{Tomoki Matsuoka}
\affiliation{Institute of Astronomy and Astrophysics, Academia Sinica, No.1, Sec.4, Roosevelt Road, Taipei 10617, Taiwan}

\author[0000-0003-2611-7269]{Keiichi Maeda}
\affiliation{Department of Astronomy, Kyoto University, Kitashirakawa-Oiwake-cho, Sakyo-ku, Kyoto 606-8502, Japan}

\author[0000-0001-5615-5464]{Yoshinori Yonekura}
\affiliation{Center for Astronomy, Ibaraki University, 2-1-1 Bunkyo, Mito, Ibaraki 310-8512, Japan}

\author[0000-0001-8537-3153]{Nozomu Tominaga}
\affiliation{Division of Science, National Astronomical Observatory of Japan, 2-21-1 Osawa, Mitaka, Tokyo 181-8588, Japan}
\affiliation{Astronomical Science Program, Graduate Institute for Advanced Studies, SOKENDAI, 2-21-1 Osawa, Mitaka, Tokyo, 181-8588, Japan}
\affiliation{Department of Physics, Faculty of Science and Engineering, Konan University, 8-9-1 Okamoto, Kobe, Hyogo, 658-8501, Japan}

\author[0000-0003-1169-1954]{Takashi J. Moriya}
\affiliation{Division of Science, National Astronomical Observatory of Japan, 2-21-1 Osawa, Mitaka, Tokyo 181-8588, Japan}
\affiliation{School of Physics and Astronomy, Faculty of Science, Monash University, Clayton, Victoria 3800, Australia}

\author[0009-0008-1070-4411]{Kenta Fujisawa}
\affiliation{The Research Institute for Time Studies, Yamaguchi University, 1677-1 Yoshida, Yamaguchi-city, Yamaguchi 753-8511, Japan}

\author[0000-0002-8169-3579]{Kotaro Niinuma}
\affiliation{The Research Institute for Time Studies, Yamaguchi University, 1677-1 Yoshida, Yamaguchi-city, Yamaguchi 753-8511, Japan}
\affiliation{Graduate School of Sciences and Technology for Innovation, Yamaguchi University, 1677-1 Yoshida, Yamaguchi-city, Yamaguchi 753-8512, Japan}

\author[0000-0002-5847-8096]{Sung-Chul Yoon}
\affiliation{Astronomy Program, Department of Physics and Astronomy, Seoul National University, Gwanak-gu, Seoul 08826, Republic of Korea}

\author[0000-0003-0894-7824]{Jae-Joon Lee}
\affiliation{Korea Astronomy and Space Science Institute, 776 Daedeokdae-ro, Yuseong-gu, Daejeon 34055, Republic of Korea}

\author[0000-0001-7003-8643]{Taehyun Jung}
\affiliation{Korea Astronomy and Space Science Institute, 776 Daedeokdae-ro, Yuseong-gu, Daejeon 34055, Republic of Korea}

\author[0000-0003-1157-4109]{Do-Young Byun}
\affiliation{Korea Astronomy and Space Science Institute, 776 Daedeokdae-ro, Yuseong-gu, Daejeon 34055, Republic of Korea}

\begin{abstract}

We report on radio follow-up observations of the nearby Type II supernova, SN 2023ixf, spanning from 1.7 to 269.9 days after the explosion, conducted using three very long baseline interferometers (VLBIs), which are the Japanese VLBI Network (JVN), the VLBI Exploration of Radio Astrometry (VERA), and the Korean VLBI Network (KVN). In three observation epochs (152.3, 206.1, and 269.9 days), we detected emission at the 6.9 and 8.4 GHz bands, with a flux density of $\sim 5$ mJy. The flux density reached a peak at around 206.1 days, which is longer than the timescale to reach the peak observed in typical Type II supernovae. Based on the analytical model of radio emission, our late-time detections were inferred to be due to the decreasing optical depth. In this case, the mass-loss rate of the progenitor is estimated to have increased from $\sim 10^{-6} - 10^{-5}\, M_{\odot}\,{\rm yr^{-1}}$ to $\sim 10^{-4}\, M_{\odot}\,{\rm yr^{-1}}$ between 28 and 6 years before the explosion. Our radio constraints are also consistent with the mass-loss rate to produce a confined circumstellar medium proposed by previous studies, which suggest that the mass-loss rate increased from $\sim 10^{-4}\, M_{\odot}\,{\rm yr^{-1}}$ to $\gtrsim 10^{-2}\, M_{\odot}\,{\rm yr^{-1}}$ in the last few years before the explosion.

\end{abstract}


\section{Introduction} \label{sec:intro}
A core-collapse supernova (CCSN) is an explosion of a massive star ($\gtrsim 8 M_{\odot}$) at its final stage of evolution. Recent advances in all-sky surveys have made it possible to conduct follow-up observations of supernovae (SNe) immediately after the explosions \citep{rau09, law09, Bellm19, Graham19}. Optical spectroscopy conducted at such a very early time has revealed the presence of narrow high-ionization emission line features (known as ``flash'' features), which originate from 
a dense circumstellar medium (CSM) in the vicinity of the supernova progenitor \citep[e.g.,][]{Khazov16,Yaron17}. Currently, a large fraction of Type II SNe is considered to have such a confined dense CSM within $\lesssim 10^{15}\,{\rm cm}$ \citep[e.g.,][]{Forster18, Bruch21, Bruch23}. This implies that their progenitor stars, which are known to be red supergiants (RSGs) for Type II SNe, underwent an enhanced mass-loss activity just decades before the explosion, with a corresponding mass-loss rate of $\gtrsim 10^{-3}\, M_{\odot}\,{\rm yr^{-1}}$. This mass-loss rate is a few orders of magnitude higher than that of the typical RSG, which is $\sim 10^{-5}\, M_{\odot}\,{\rm yr^{-1}}$ \citep[e.g.,][]{Goldman17}.

The radio emission of CCSNe is interpreted to be the synchrotron emission of relativistic electrons that are accelerated by the SN-CSM interaction \citep{Chevalier06a,Chevalier06b}. This makes it a unique tool for probing the CSM density structure and mass-loss history of the progenitor star. However, whereas thousands of SNe have been detected in the optical band, only a few hundred have been observed in radio wavelengths, even when undetected events are taken into account \citep{Bietenholz21}. Only a small fraction of radio SNe have been observed with high cadence and/or in multiple radio frequency bands. It is known from the light curves of extensively observed samples that many of them exhibit complex fluctuations that cannot be explained by simple models that predict monotonic increase and decrease in the expected luminosity evolution (e.g., SN 1979C, \citealt{Weiler92}; SN 2001em, \citealt{Chandra20}; SN 2001ig, \citealt{Ryder04}; SN 2003bg, \citealt{Soderberg06}, SN 2004dk \citealt{Balasubramanian21}, SN 2014C \citealt{Anderson17}, and SN 2018ivc, \citealt{Maeda23}). Some cases may be associated with the complexity of the CSM density structure.

SN 2023ixf was discovered on 2023 May 19.727 UT, ${\rm MJD}=60083.727$ \citep{Itagaki23}, in the nearby galaxy M101, at a distance of $6.85 \pm 0.15$ Mpc \citep{Riess22}, and classified as a Type II SN \citep{Perley23}. The time of the SN first light is estimated to be ${\rm MJD} = 60082.743 \pm 0.083$ \citep{Hiramatsu23}. Because it is the closest CCSN in a decade, extensive electromagnetic follow-up observations of SN 2023ixf were conducted across a wide range of wavelengths. 
The early excess in the optical light curve \citep{Jacobson-Galan23, Hosseinzadeh23, Sgro23}, the flash features in the optical spectra \citep{Yamanaka23, Jacobson-Galan23, Smith23, Bostroem23, Teja23}, and early X-ray detections \citep{Grefenstette23, Chandra24} suggest the presence of a dense CSM and the interaction of SN ejecta with it, but there is an apparent inconsistency between the mass-loss rates estimated by optical \citep[$\sim 10^{-2}\, M_{\odot}\,{\rm yr^{-1}}$, ][]{Jacobson-Galan23} and X-ray \citep[$\sim 10^{-4}\, M_{\odot}\,{\rm yr^{-1}}$,][]{Grefenstette23} observations. The dense and asymmetric CSM, suggested by a change in its position angle observed by early-time spectropolarimetry \citep{Vasylyev23}, is considered to be the cause of this discrepancy.

The progenitor candidate of SN 2023ixf has been identified in the archival images of the Spitzer Space Telescope and Hubble Space Telescope \citep{Szalai23, Pledger23}. By intensive archival data analyses, the progenitor candidate is identified as a luminous ($\sim 10^5\, L_{\odot}$), dust-obscured RSG, exhibiting a possible periodic variability with a period of $\sim 1000$ days \citep{Jencson23, Kilpatrick23, Soraisam23ApJ, VanDyk24, Niu23, Neustadt24}. The initial mass of the progenitor is, however, controversial, with different values obtained through different methods (e.g., based on luminosity, variability, and stellar populations in the vicinity of SN 2023ixf). The observed progenitor candidate activity disfavors the presence of any outburst events in the last $\sim 20$ years \citep{Jencson23, Dong23, Neustadt24}.

Although the observational results on SN 2023ixf suggest the presence of a confined dense CSM, the radial density profile of the CSM, including the extended tenuous region, has not been tightly constrained. Radio observations provide information on the SN-CSM interaction and make it possible to estimate the mass-loss rate of the progenitor star independently from the other wavelength observations. Regarding the radio observations of SN 2023ixf, \citet{Berger23} reported non-detection at 230 GHz during the early time ($\lesssim 18$ days after the explosion). Observations at other radio wavelengths and with longer time spans will be beneficial to elucidate more details on the progenitor evolution. In this paper, we present the results of radio follow-up observations of SN 2023ixf in the frequency range from 6 GHz to 129 GHz by three Japanese and Korean very long baseline interferometers (VLBIs): the Japanese VLBI Network (JVN), the VLBI Exploration of Radio Astrometry (VERA), and the Korean VLBI Network (KVN). VLBI observation for SNe is usually conducted to resolve the spatial structure of nearby objects, but SN 2023ixf is expected to be unresolved at this stage so we used them only to measure the flux densities. The VLBI arrays of our observations consist of fewer antennas than those in recent large-scale VLBIs but are superior in their capability for rapid and flexible scheduling and their use of optimized characteristics for each array. We describe radio observations of SN 2023ixf in Section \ref{sec:obs}, report measured flux densities in Section \ref{sec:result}, discuss the mass-loss history of the progenitor star in Section \ref{sec:discuss}, and summarize our findings in Section \ref{sec:conclusion}.

\section{Observations and Data Analysis} \label{sec:obs}
\subsection{JVN}
\label{subsec:jvn}
The JVN observations were conducted as a single baseline VLBI using two radio telescopes: Hitachi 32-m (Hit32) \citep{Yonekura16} and Yamaguchi 34-m (Yam34) \citep{Fujisawa22}. The details of the observations are summarized in Table \ref{tab:jvn}. The observations were conducted in 12 epochs spanning from 1 day to 270 days after the discovery report on SN 2023ixf. The observation intervals were set to be roughly equal on the logarithmic scale. Simultaneous observations were made on two frequency bands, one centered on 6.856 GHz (C band) and the other on 8.448 GHz (X band), each with a bandwidth of 512 MHz. Left-hand circular polarizations were received and sampled with 2-bit quantization. The recording speed was 2048 Mbps for each frequency band. The distance between Hit32 and Yam34 is 873 km, corresponding to angular resolutions of 10 mas at the C band and 8.4 mas at the X band. We observed the quasar 3C345 as a fringe finder and ICRF radio source J135905.7+554429 (J1359+5544) as a gain calibrator. The on-source integration times for 3C345, J1359+5544, and SN 2023ixf were 10 min, 4 min, and 9 min, respectively. The target and gain calibrator observations were repeated three times.

The JVN data were correlated using the FX-type software correlator GICO3, developed by the National Institute of Information and Communications Technology. We first measured a clock delay and rate by a fringe search for 3C345. Adopting the delay and rate, we performed fringe searches for J1359+5544 and SN 2023ixf, and then obtained the correlated amplitude and signal-to-noise ratio of each target. Using the correlated amplitude of 3C345 observed with JVN and the absolute flux density value of 3C345 measured by the single-dish observations of Hit32, we determined a flux density scaling corresponding to the given correlated amplitude. In the flux scaling, we also corrected the aperture efficiency of Hit32 for its dependence on the elevation angle according to \citet{Yonekura16}. For Yam34 we used the average aperture efficiency of 0.65 \citep{Fujisawa22}. The spatial structure of 3C345 does not significantly affect the flux calibration because the $uv$-distance between Hit32 and Yam34 corresponds to 20 M$\lambda$ at the C band and the correlated amplitude of 3C345 hardly decreases at this distance\footnote{see \url{https://obs.vlba.nrao.edu/cst/calibsource/12470}.}. In this single baseline VLBI observation, it is not possible to produce an image, and detection or non-detection is based on the examination of the fringe peak. We report as a detection if a $> 3\sigma$ peak is located close to the delay and rate estimated by the gain calibrator consistently throughout three observations in one epoch (delay within $\pm 10$ ns and rate within $\pm 15$ mHz), and conservatively adopt $7\sigma$ as an upper limit on non-detections. Here, we took the $1\sigma$ uncertainty to be the standard deviation of the correlated amplitude in the fringe space. We then obtained the flux density by averaging the flux densities (or 1$\sigma$ uncertainty in the case of non-detection) of three scans in each epoch, based on the assumption that the flux density does not vary within the $\sim 30$ mim of observations of the three scans. We note that the scatter (standard deviation) of the detected three flux densities in each epoch was $\lesssim 1\,{\rm mJy}$, comparable to or less than the 1$\sigma$ levels derived from the correlated amplitude. To estimate the uncertainty of the detected flux density, we used the averaged 1$\sigma$ correlated amplitude uncertainty in three scans ($\sigma_{\rm CA}$), and the scatter of the three scans ($\sigma_{\rm Sc}$), as $\sqrt{(\sigma_{\rm CA})^2 + (\sigma_{\rm Sc})^2}$. We also evaluated the uncertainty of the flux density variation of the flux calibrator 3C345, as 12\% for both the C and X band from the flux density monitoring data of 3C345 since 2020 June by Hit32. Since we used the same flux density value of 3C345 for the scaling of all the epochs, it only changes the absolute flux density scaling and does not affect the relative variation of the measured correlated amplitude. There are other uncertainties due to the pointing accuracy and due to the chopper wheel method. However, these can be ignored, because the former is estimated to be $\le 1$\% based on the pointing accuracy of $\le 0.3\arcmin$ for Hit32 \citep{Yonekura16} and $\le 1\arcmin$ for Yam34 \citep{Fujisawa22}, and the latter originating from the difference in the actual temperatures between the atmosphere and the black body can be calculated to be $< 1$\%. Therefore, for the JVN-detected flux density, we used the averaged flux density and error derived from the three scans (shown as the averaged ones in Table \ref{tab:jvn}).

\subsection{VERA}
The VERA observations were conducted in the Director's Discretionary Time (DDT) of the VERA Large-scale COllaborative Program (VLCOP). Three of four stations of VERA, i.e., Mizusawa, Ogasawara, and Iriki 20-m telescopes, participated in the observations. Two frequency bands, centered at 22.484 GHz (K band) and 43.283 GHz (Q band), were observed in 3 epochs respectively. The phase calibrator J1359+5544 and SN 2023ixf were observed simultaneously using the dual-beam mode for the phase referencing. We also observed the quasar 3C273 as a bandpass and delay calibrator. Left-hand circular polarization with a total bandwidth of 2048 MHz was recorded in the 16 Gbps recording mode (8 Gbps for the target and 8 Gbps for the phase calibrator) using a new digital backend system consisting of an OCTAve A/D converter (OCTAD) sampler and an OCTADISK2 recorder. The details of the newly equipped VERA observing system are described in \citet{Oyama24}. Our VERA observations are summarized in Table \ref{tab:vera}.

The data were correlated using a software FX-type correlator, OCTACOR2, installed at the Mizusawa VLBI Observatory. Data reduction was performed using the Astronomical Image Processing System (AIPS) developed at the National Radio Astronomy Observatory (NRAO). The amplitude was calibrated based on the system noise temperatures, and bandpass calibration was made using the 3C273 data. The modified delay-tracking model was applied for accurate measurements. We produced the image of J1359+5544 via the task CLEAN and used it for the dual-beam phase calibration. The flux densities of J1359+5544 at the K band and Q band were $\sim 300\,{\rm mJy}$ and $100-200\,{\rm mJy}$, respectively. Finally, we obtained the images around SN 2023ixf. We inspected the images and set the upper limit at $5\sigma$ in the case of non-detection.

\subsection{KVN}
The KVN observations were conducted using three KVN antennas: Yonsei (KY), Ulsan (KU), and Tamna (KT) 21-m radio telescopes. The observations are summarized in Table \ref{tab:kvn}. We note that KT did not join in the observation of t23sy01i. The longest baseline length is 476 km, between KY and KT. We observed four frequency bands, which are centered at 22.062 (K band), 43.612 (Q band), 86.712 (W band), and 129.812 GHz (D band), via the simultaneous multi-frequency VLBI System \citep{Han13}. Each frequency band consists of 2 channels $\times$ 512 MHz bandwidth. The phase calibrator, J1419+5423, which is 2\fdg 35 away from SN 2023ixf, was observed using the fast-switching mode. For bandpass and delay calibrators, we selected 3C345, 3C279, 0836+710 (also known as 4C 71.07), and J0927+3902 (also known as 4C 39.25), depending on the observed elevation angle. Left-hand circular polarization with a $4\times 1024$ MHz bandwidth was recorded in the 16 Gbps recording mode using the recording system of Mark6 or Flexbuff. Data correlation was performed using the DiFX software correlator in the Korea–Japan Correlator Center.

Data reduction was performed using AIPS, in the same manner as that done on the VERA data, i.e., the amplitude calibration was performed based on the system noise temperature. Since the phase calibrator was partially detected in the W band and not detected in the D band, we analyzed only the K and Q bands data. We inspected the images produced with the Difmap software and set the upper limit at $5\sigma$ in the case of non-detection. 

\begin{deluxetable*}{cccccccc}
\tablecaption{JVN observations\label{tab:jvn}}
\tablewidth{0pt}
\tablehead{
\colhead{Obscode} & \colhead{Date} & \colhead{MJD} & \colhead{$t_{\rm exp}$} & \colhead{$F_{\rm 6.9\,GHz}$} & \colhead{$F_{\rm 8.4\,GHz}$} & \colhead{$L_{\rm 6.9\,GHz}$} & \colhead{$L_{\rm 8.4\,GHz}$}\\
\colhead{} & \colhead{(UT)} & \colhead{} & \colhead{(days)} &
\multicolumn{2}{c}{(mJy)} & \multicolumn{2}{c}{($10^{26}\,{\rm erg\,s^{-1}\,Hz^{-1}}$)}
}
\startdata
U23140A & 2023-05-20 10:39:30 & 60084.44 & 1.70 & $<$4.7 & $<$5.5 & $<$2.7 & $<$3.1 \\
U23144A & 2023-05-24 09:53:30 & 60088.41 & 5.67 & $<$4.8 & $<$5.4 & $<$2.7 & $<$3.0 \\
U23147A & 2023-05-27 09:41:30 & 60091.40 & 8.66 & $<$5.5 & $<$6.5 & $<$3.1 & $<$3.6 \\
U23157A & 2023-06-06 09:02:30 & 60101.38 & 18.63 & $<$6.0 & $<$6.5 & $<$3.4 & $<$3.7 \\
U23168A & 2023-06-17 08:18:30 & 60112.35 & 29.60 & $<$5.3 & $<$6.4 & $<$3.0 & $<$3.6 \\
U23185A & 2023-07-04 07:12:30 & 60129.30 & 46.56 & $<$6.5 & $<$8.0 & $<$3.7 & $<$4.5 \\
U23200A & 2023-07-19 06:13:30 & 60144.26 & 61.52 & $<$17.1 & $<$17.4 & $<$9.6 & $<$9.8 \\
U23217A & 2023-08-05 05:05:30 & 60161.21 & 78.47 & $<$13.5 & $<$19.1 & $<$7.6 & $<$10.7 \\
U23247A & 2023-09-04 03:07:30 & 60191.13 & 108.39 & $<$36.2 & $<$70.2 & $<$20.3 & $<$39.4 \\\hline
U23290A & 2023-10-17 23:59:30 & 60235.00 & 152.26 & $5.0\pm0.7$ & $3.8\pm0.8$ & $2.8\pm0.4$ & $2.2\pm0.5$ \\
 & 2023-10-18 00:14:30 & 60235.01 & 152.27 & $3.4\pm0.8$ & $3.3\pm0.8$ & $1.9\pm0.4$ & $1.8\pm0.5$ \\
 & 2023-10-18 00:29:30 & 60235.02 & 152.28 & $2.8\pm0.8$ & $3.8\pm0.8$ & $1.6\pm0.4$ & $2.1\pm0.5$ \\
--average & 2023-10-18 00:14:30 & 60235.01 & 152.27 & $3.8\pm1.0$ & $3.6\pm0.5$ & $2.1\pm0.6$ & $2.0\pm0.6$ \\\hline
U23344A & 2023-12-10 20:28:30 & 60288.85 & 206.11 & $5.9\pm0.8$ & $6.6\pm1.0$ & $3.3\pm0.4$ & $3.7\pm0.6$ \\
 & 2023-12-10 20:43:30 & 60288.86 & 206.12 & $5.9\pm0.8$ & $7.4\pm1.0$ & $3.3\pm0.5$ & $4.2\pm0.6$ \\
 & 2023-12-10 20:58:30 & 60288.87 & 206.13 & $5.8\pm0.8$ & $4.8\pm1.0$ & $3.3\pm0.4$ & $2.7\pm0.6$ \\
--average & 2023-12-10 20:43:30 & 60288.86 & 206.12 & $5.9\pm0.5$ & $6.3\pm1.2$ & $3.3\pm0.3$ & $3.5\pm0.3$ \\\hline
U24043A & 2024-02-12 16:16:30 & 60352.68 & 269.94 & $4.3\pm0.6$ & $3.3\pm0.8$ & $2.4\pm0.4$ & $1.8\pm0.5$ \\
 & 2024-02-12 16:31:30 & 60352.69 & 269.95 & $4.2\pm0.6$ & $3.3\pm0.8$ & $2.4\pm0.4$ & $1.8\pm0.5$ \\
 & 2024-02-12 16:46:30 & 60352.70 & 269.96 & $4.5\pm0.7$ & $1.9\pm0.8$ & $2.5\pm0.4$ & $1.1\pm0.5$ \\
--average & 2024-02-12 16:31:30 & 60352.69 & 269.95 & $4.4\pm0.4$ & $2.8\pm0.8$ & $2.4\pm0.2$ & $1.6\pm0.2$ \\
\enddata
\tablecomments{Dates are at the midpoints of each observation. Time since explosion ($t_{\rm exp}$) is given relative to an estimated explosion time of MJD = 60082.743 \citep{Hiramatsu23}. The integration time of each observation is 9 min. The upper limits of the flux density correspond to 7$\sigma$ of the correlated amplitude. The flux density errors of the three scans of each detected epoch correspond to the 1$\sigma$ correlated amplitude uncertainties. The averaged flux density error was obtained by combining the averaged 1$\sigma$ correlated amplitude uncertainty and the scatter of the three scans (see Section \ref{subsec:jvn}).
}
\end{deluxetable*}

\begin{deluxetable*}{cccccccc}
\tablecaption{VERA observations\label{tab:vera}}
\tablewidth{0pt}
\tablehead{
\colhead{Obscode} & \colhead{Date} & \colhead{MJD} & \colhead{$t_{\rm exp}$} & \colhead{$T_{\rm integ}$}& \colhead{Frequency} & \colhead{$F_\nu$} & \colhead{$L_{\nu}$}\\
\colhead{} & \colhead{(UT)} & \colhead{} & \colhead{(days)} & \colhead{(min)} & \colhead{(GHz)} & \colhead{(mJy)} & \colhead{($10^{26}\,{\rm erg\,s^{-1}\,Hz^{-1}}$)} 
}
\startdata
R23145A & 2023-05-25 11:56:30 & 60089.50 & 6.75 & 360 & 43.283 & $<$2.3 & $<$1.3 \\
R23147A & 2023-05-27 09:38:30 & 60091.40 & 8.66 & 120 & 22.484 & $<$2.8 & $<$1.6 \\
R23148A & 2023-05-28 11:50:30 & 60092.49 & 9.75 & 360 & 43.283 & $<$2.9 & $<$1.6 \\
R23151A & 2023-05-31 12:00:30 & 60095.50 & 12.76 & 360 & 22.484 & $<$1.8 & $<$1.0 \\
R23152A & 2023-06-01 11:30:30 & 60096.48 & 13.74 & 360 & 43.283 & $<$4.0 & $<$2.2 \\
R23157A & 2023-06-06 08:58:30 & 60101.37 & 18.63 & 120 & 22.484 & $<$4.7 & $<$2.6 \\
\enddata
\tablecomments{Dates and $t_{\rm exp}$ are obtained in the same manner as Table \ref{tab:jvn}. The integration time ($T_{\rm integ}$) represents the total on-source time of each epoch. The upper limits of the flux density correspond to 5$\sigma$.}
\end{deluxetable*}

\begin{deluxetable*}{ccccccccc}
\tablecaption{KVN observations\label{tab:kvn}}
\tablewidth{0pt}
\tablehead{
\colhead{Obscode} & \colhead{Date} & \colhead{MJD} & \colhead{$t_{\rm exp}$} & \colhead{$T_{\rm integ}$}& \colhead{$F_{\rm 22\,GHz}$} & \colhead{$F_{\rm 43\,GHz}$} & \colhead{$L_{\rm 22\,GHz}$} & \colhead{$L_{\rm 43\,GHz}$}\\
\colhead{} & \colhead{(UT)} & \colhead{} & \colhead{(days)} & \colhead{(min)} &
\multicolumn{2}{c}{(mJy)} & \multicolumn{2}{c}{($10^{26}\,{\rm erg\,s^{-1}\,Hz^{-1}}$)}
}
\startdata
t23sy01i & 2023-06-12 14:22:20 & 60107.60 & 24.86 & 75.0 & $<$6.0 & $<$5.2 & $<$3.4 & $<$3.4 \\
t23sy01j & 2023-06-15 15:42:39 & 60110.65 & 27.91 & 56.2 & $<$3.6 & $<$4.2 & $<$2.1 & $<$2.1 \\
t23sy01k & 2023-06-18 14:42:16 & 60113.61 & 30.87 & 56.2 & $<$3.4 & $<$2.8 & $<$1.9 & $<$1.9 \\
\enddata
\tablecomments{Dates and $t_{\rm exp}$ are obtained in the same manner as Table \ref{tab:jvn}. The integration time ($T_{\rm integ}$) represents the total on-source time of each epoch. The upper limits of the flux density correspond to 5$\sigma$.}
\end{deluxetable*}

\section{Results} \label{sec:result}
\begin{figure}
\plotone{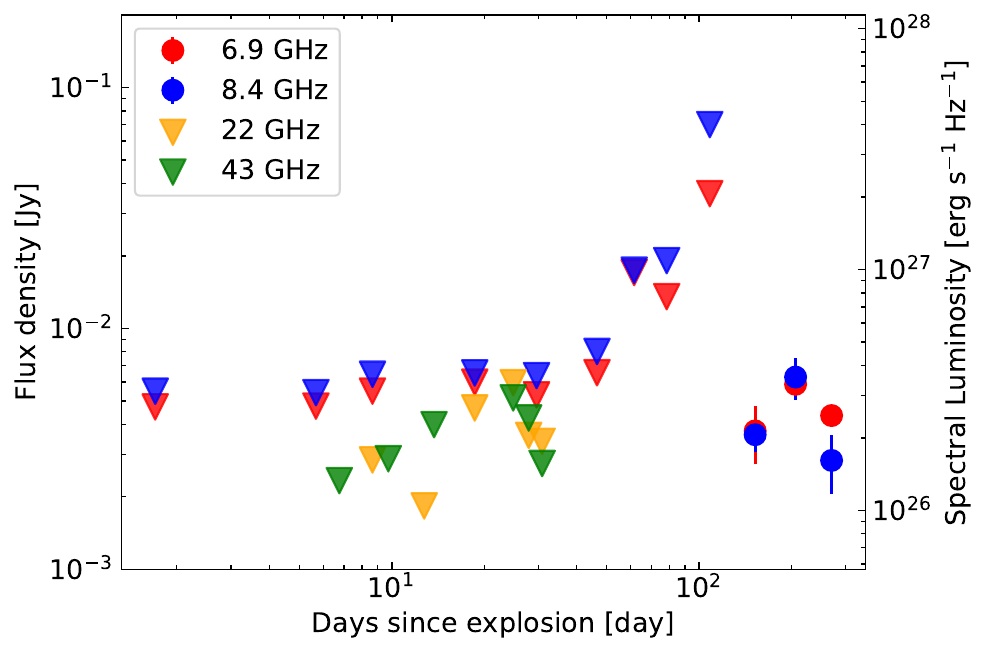}
\caption{Observed flux densities and spectral luminosities of SN 2023ixf with JVN (6.9 and 8.4 GHz), VERA (22 and 43 GHz), and KVN (22 and 43 GHz). The upper limits of JVN are 7$\sigma$, while those of VERA and KVN are 5$\sigma$. Circles and triangles indicate detections and non-detections, respectively.
\label{fig:alldata}}
\end{figure}

The measured flux densities of SN 2023ixf are summarized in Tables \ref{tab:jvn}, \ref{tab:vera} and \ref{tab:kvn}, and shown in Figure \ref{fig:alldata}. The time since explosion ($t_{\rm exp}$) of each observation was calculated using the estimated SN first light of MJD = 60082.743 \citep{Hiramatsu23}. The JVN observation epochs at the C/X bands covered a wide range for the time span with a high cadence. The elevated noise levels at the C/X bands observed in U23200A and U23217A were due to problems with the Yam34 receiver, and those in U23247A were attributed to severe weather conditions at the Hitachi 32-m site. With the exception of those three observations, we achieved a sensitivity of a few mJy. There has been a reported detection of $40\,{\rm \mu Jy}$ using the Very Large Array (VLA) in 10 GHz at $t_{\rm exp}$ of 29.28 days \citep{Matthews23}, and this is consistent with our non-detections at the mJy level. Comparing our first detection with the VLA detection, we inferred that SN 2023ixf became two orders of magnitude brighter in 120 days.

\begin{figure}
\plotone{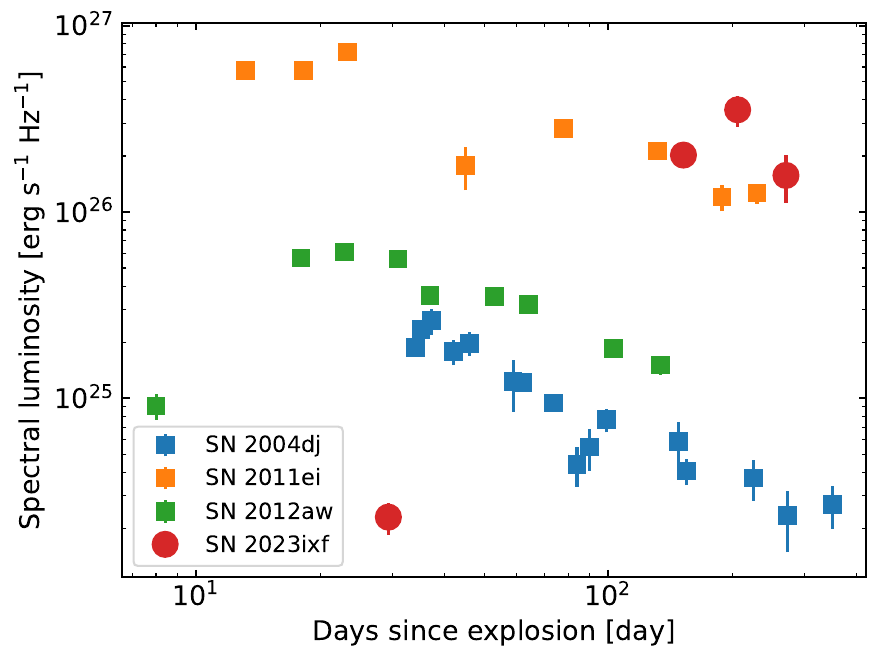}
\caption{Spectral luminosity evolution of SN 2023ixf compared with those of other Type II SNe (blue: SN 2004dj from \citealt{Nayana18}, orange: SN 2011ei from \citealt{Milisavljevic13}, green: SN 2012aw from \citealt{Yadav14}). All measurements were in the 8 GHz band. The first point for SN 2023ixf was measured with the VLA \citep{Matthews23}, whereas the latter three are based on our JVN data.
\label{fig:comparison}}
\end{figure}

The detected spectral luminosity of a few $10^{26} \,{\rm erg\,s^{-1}\, Hz^{-1}}$ is higher than the mean peak spectral luminosity of $L_{\rm pk} = 10^{25.3\pm 1.3}\,{\rm erg\,s^{-1}\, Hz^{-1}}$ for Type II SNe \citep{Bietenholz21} but not unusual considering the large range of values seen in other SNe. Figure \ref{fig:comparison} shows the luminosity evolution of SN 2023ixf in the 8.4 GHz band in comparison to those of three other Type II SNe: SN 2004dj, SN 2012aw, and SN 2011ei. We note that the estimated mass-loss rates are $\sim 10^{-6}\, M_{\odot}\,{\rm yr^{-1}}$ for SN 2004dj and SN 2012aw \citep{Nayana18, Yadav14}, and $\sim 10^{-5}\, M_{\odot}\,{\rm yr^{-1}}$ for SN 2011ei \citep{Milisavljevic13}. The radio flux density of SN 2023ixf seems to have reached a peak at around 206 days. This timescale to reach the peak ($t_{\rm pk}$) is relatively long compared to the typical value of $10^{1.6\pm1.0}$ days for Type II SNe \citep{Bietenholz21} but still within the range of the variation. The higher luminosity and later peak of SN 2023ixf imply the presence of dense CSM. 

Comparing $L_{\rm pk}$ and $t_{\rm pk}$ of SN 2023ixf with those of normal type II SNe obtained in the 4 GHz to 10 GHz frequency range presented in \citet{Bietenholz21}, we found that 2 of 23 SNe have characteristics similar to those of SN 2023ixf, $L_{\rm pk} > 10^{26.5}\,{\rm erg\,s^{-1}\, Hz^{-1}}$ and $t_{\rm pk} > 150$ days (see Figure 10 in \citealt{Bietenholz21}). One of the similar SNe is 1979C, which has an estimated mass-loss rate of $\sim 10^{-4}\, M_{\odot}\,{\rm yr^{-1}}$ \citep{Lundqvist88, Weiler91}. The early detection of such late-bright SNe is difficult in the case of distant SNe. For example, if SN 2023ixf were at 15 Mpc, its flux density at 29 days ($40\,{\rm \mu Jy}$ at 6.85 Mpc) would be 8 $\mu$Jy, which is hard to detect. If it is assumed that the spectral luminosity at 100 days is $10^{26}\,{\rm erg\, s^{-1}\, Hz^{-1}}$, objects within 50 Mpc can be detected with a sensitivity of tens of $\mu$Jy, and only objects within 5 Mpc can be detected with sensitivity at mJy level. Thus, many SN 2023ixf-like radio light curves would have been missed thus far if a follow-up project terminates in $\lesssim 100$ days because of the non-detection.

\section{Discussion} \label{sec:discuss}
\subsection{Constraints on the Mass-loss Rate of the Progenitor}\label{subsec:mdot_constraints}

Using the measured flux densities and upper limits, we infer the CSM structure of SN 2023ixf, which can be translated into the mass-loss history of the progenitor star by assuming the wind velocity. Here, we obtained the lower limits of the mass-loss rate considering the optically thick limit assuming free-free absorption as a dominant absorption process following \citet{Weiler86}, and the upper limits considering the optically thin synchrotron emission model by \citet{Chevalier98}. Those assumptions are the same as those adopted by \citet{Berger23}, but the specific models are different from theirs.

Our assumptions are as follows. The cooling process is dominated by an adiabatic expansion. The energy spectral index of relativistic electrons ($p$) is 3 \citep{Chevalier06b, Maeda12}. The minimum Lorentz factor of the accelerated electrons ($\gamma_{\rm min}$) is 1. We employed the self-similar solution of \citet{Chevalier82} to calculate the time evolution of shock velocity. We assume the outer density profile of the expanding ejecta as $\rho_{\rm ej} \propto r^{-12}$ \citep{Matzner99} and a CSM density profile of $\rho_{\rm CSM} \propto r^{-2}$.

In the free–free absorption model, the optical depth $\tau_{\rm FFA}$ is given by
\begin{eqnarray}
\tau_{\rm FFA} \sim && 0.9 \left(\frac{T_e}{10^4\,{\rm K}}\right)^{-3/2}  \left(\frac{\dot{M}}{10^{-6}\,M_{\odot}\,{\rm yr^{-1}}}\right)^{23/10} \nonumber \\
&&\times \left(\frac{v_{\rm w}}{115\,{\rm km\,s^{-1}}}\right)^{-23/10} \left(\frac{E_{\rm kin}}{10^{51}\,{\rm erg}}\right)^{-27/20}\nonumber\\
&&\times \left(\frac{M_{\rm ej}}{10\,M_{\odot}}\right)^{21/20} \left(\frac{t}{10\,{\rm day}}\right)^{-27/10} \left(\frac{\nu}{1\,{\rm GHz}}\right)^{-2}.
\label{eq:ffa}
\end{eqnarray}
By assuming that the non-detection was due to $\tau_{\rm FFA} \ge 1$, we can obtain the lower limits of the mass-loss rate ($\dot{M}$) of the progenitor. The typical value of electron temperature in an unshocked CSM ($T_e$) is $\sim 10^{4-5}\,{\rm K}$ \citep[see][]{Lundqvist88, Chevalier06a}, and we used the conservative one of $10^4\,{\rm K}$ because the higher $T_e$ gives a tighter constraint. Using a wind velocity of $v_{\rm w} = 115\,{\rm km\,s^{-1}}$ \citep{Smith23}, kinetic energy of $E_{\rm kin} = 1.2 \times 10^{51}\,{\rm erg}$ and ejecta mass of $M_{\rm ej} = 9.4\, M_{\odot}$ \citep{Bersten24}, we calculated the lower limits of the mass-loss rate at each time and each frequency band from the non-detection data. 

The flux density $F_\nu$ of the optically thin synchrotron emission can be described to be
\begin{eqnarray}
F_{\nu} =&& 4.49\times 10^{2}\,{\rm mJy}\, \left(\frac{\dot{M}}{10^{-6}\,M_{\odot}\,{\rm yr^{-1}}}\right)^{29/20} \nonumber \\
&&\times \left(\frac{\epsilon_e}{0.1}\right) \left(\frac{\epsilon_B}{0.1}\right)^{4/3} \left(\frac{v_{\rm w}}{115\,{\rm km\,s^{-1}}}\right)^{-29/20} \nonumber \\
&&\times \left(\frac{E_{\rm kin}}{10^{51}\,{\rm erg}}\right)^{27/20} \left(\frac{M_{\rm ej}}{10\,M_{\odot}}\right)^{-21/20}  \nonumber \\
&&\times \left(\frac{D}{6.85\,{\rm Mpc}}\right)^{-2}\left(\frac{t}{10\,{\rm day}}\right)^{-4/5} \left(\frac{\nu}{1\,{\rm GHz}}\right)^{-1},
\label{eq:ssa}
\end{eqnarray}
where $\epsilon_e$ and $\epsilon_B$ are the postshock energy fractions in the relativistic electron and magnetic fields, respectively, and $D$ is the distance to the source. Note that this model depends highly on the uncertain parameters of $\epsilon_e$ and $\epsilon_B$, and we used the conventional values of $\epsilon_e = \epsilon_B = 0.1$.

\begin{figure*}
\plotone{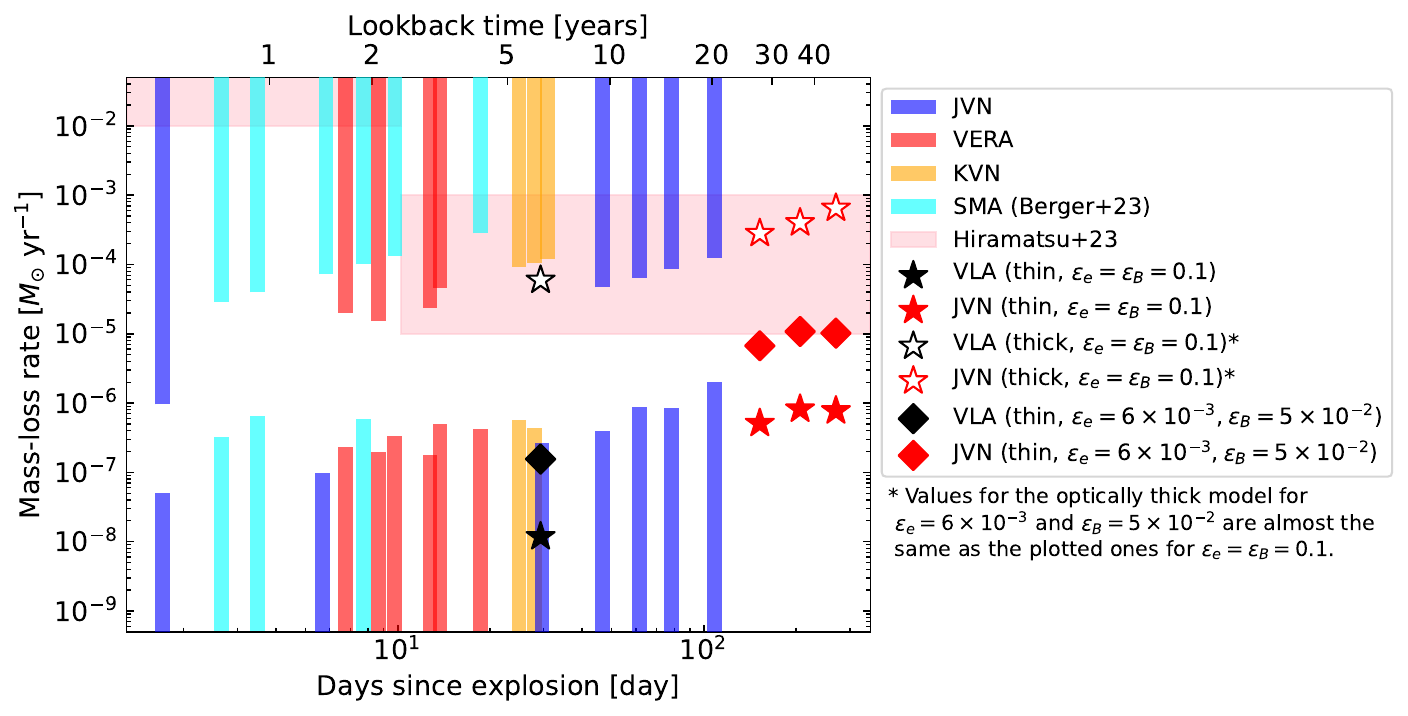}
\caption{Constraints on progenitor mass-loss rate of SN 2023ixf. The allowed mass-loss rate regions from the free-free absorption model are shown by upper vertical lines, and those from the optically thin synchrotron emission model are shown by the lower vertical lines, by calculating with parameters of $T_e = 10^4\,{\rm K}$, $M_{\rm ej} = 9.4\, M_{\odot}$, $E_{\rm kin} = 1.2 \times 10^{51}\,{\rm erg}$, $v_{\rm w} = 115\,{\rm km\,s^{-1}}$, $\epsilon_e = \epsilon_B = 0.1$. Stars represent the mass-loss rate derived from flux densities detected with the VLA \citep{Matthews23} and JVN, using the optically thin (color-filled) and optically thick models (unfilled). Diamonds are the mass-loss values calculated with the optically thin model for $\epsilon_e = 6\times 10^{-3}$ and $\epsilon_B = 5\times 10^{-2}$. Values for the optically thick model for $\epsilon_e = 6\times 10^{-3}$ and $\epsilon_B = 5\times 10^{-2}$ are almost the same as the plotted ones for $\epsilon_e = \epsilon_B = 0.1$. The pink-shaded regions represent the ranges of mass-loss rate proposed by \citet{Hiramatsu23}. For clarity, we plotted only the tightest constraint when different regions overlapped owing to multi-frequency observations or the same observing date. The lookback time on the upper x-axis was calculated by assuming a shock velocity at 10 days after the explosion of $10^4\,{\rm km\,s^{-1}}$, shock deceleration of $V \propto (t/{\rm 10\, days})^{-0.1}$, and constant wind velocity of $115\,{\rm km\,s^{-1}}$.
\label{fig:mdot}}
\end{figure*}

Figure \ref{fig:mdot} shows the constraints on the mass-loss rate history of the progenitor star of SN 2023ixf, calculated using our measured flux densities combined with the Submillimeter Array (SMA) 230 GHz upper limits \citep{Berger23}. The vertical lines in Figure \ref{fig:mdot} indicate the permitted mass-loss rate either in the optically thick limit (to free-free absorption) or in the optically thin limit. Our constraints exclude some regions ranging from $\sim 10^{-5}$ to $\sim 10^{-3}\, M_{\odot}\,{\rm yr^{-1}}$. The $t_{\rm exp}$ from 1.7 days to 269.9 days corresponds to the radius from $1.4\times 10^{14}\, {\rm cm}$ to $1.3\times 10^{16}\, {\rm cm}$, which is obtained by assuming a shock velocity at 10 days after the explosion of $10^4\,{\rm km\,s^{-1}}$ and a shock deceleration of $V \propto (t/{\rm 10\, days})^{-0.1}$. Moreover, assuming a constant wind velocity of $115\,{\rm km\,s^{-1}}$, we can estimate the lookback time from the explosion to be 46 years.

Many past studies on SN 2023ixf have proposed a model of a CSM confined to the vicinity of the progenitor star \citep[e.g.,][]{Jacobson-Galan23, Teja23, Hiramatsu23}. For example, \citet{Hiramatsu23} proposed $\dot{M} \ge 10^{-2}\, M_{\odot}\,{\rm yr^{-1}}$ within $R \lesssim 7\times 10^{14}\,{\rm cm}$ and orders of $10^{-5}$ to $10^{-4}\, M_{\odot}\,{\rm yr^{-1}}$ at a larger radius, shown as the pink-shaded region in Figure \ref{fig:mdot}. A mass-loss rate of several times $10^{-4}\, M_{\odot}\,{\rm yr^{-1}}$ is also suggested by the X-ray observations \citep{Grefenstette23, Chandra24}. Whereas the high mass-loss rate in the early $t_{\rm exp}$ is consistent with our constraints, the inferred mass-loss rate lower than $3\times 10^{-4}\, M_\odot\,{\rm yr}^{-1}$ at 19 days after the explosion is not consistent with our measurement given the assumed parameter values and model.

Indeed, radio SNe that brighten at a later time ($t_{\rm exp} \sim 1000$ days) may provide insight into the origin of the possible increase in mass loss of SN 2023ixf. Late radio brightening is typically observed in Type IIn SNe and interpreted as an outcome of a high-density CSM \citep[e.g.,][]{Chandra15}. The late radio brightening of SN 2023ixf therefore suggests that it has the properties of a Type II SN in an environment which is dense, but not as extremely dense as those of Type IIn SNe.

From the detected flux densities obtained by the VLA and JVN, we can estimate the corresponding mass-loss rates. Here, we present two possible interpretations, one assuming the optically thin limit (Section \ref{subsec:thin}) and the other including the optically thick effect to both free-free absorption and synchrotron self-absorption (Section \ref{subsec:thick}). We then discuss the mass-loss history of the progenitor star (Section \ref{subsec:mass-loss}).

\subsection{Optically Thin Interpretation \label{subsec:thin}}
Using the optically thin synchrotron emission model (also optically thin to free-free absorption) as shown in Equation (\ref{eq:ssa}), we obtained the mass-loss rate corresponding to each detection, shown as the black and red filled symbols in Figure \ref{fig:mdot}. This indicates that the mass-loss rate had decreased by approximately two orders of magnitude somewhere between 28 years before the explosion (JVN detection at 152 days, $1\times 10^{16}\,{\rm cm}$) and 6 years before the explosion (VLA detection at 29 days, $2\times 10^{15}\,{\rm cm}$), if we assume that all the VLA- and JVN-detected emissions are optically thin. Note that the absolute values of the mass-loss rates depend on $\epsilon_e$ and $\epsilon_B$. However, the relative decrease in the mass-loss rate obtained from the JVN observations with respect to that obtained from the VLA is independent of these parameters (assuming both observations were in the optically thin stage). We note that the mass-loss rates estimated from the analysis of the optically thin assumption satisfy $\tau_{\rm SSA} < 1$, where $\tau_{\rm SSA}$ is an optical depth of the internal synchrotron self-absorption, given by Equation (\ref{eq:tau_ssa}) in Section \ref{subsec:thick}.

\subsection{Optically Thick Interpretation \label{subsec:thick}}
For the opposite case, we derive the mass-loss rates under the assumption of optically thick for either or both free-free absorption and synchrotron self-absorption. The optical depth of the internal synchrotron self-absorption ($\tau_{\rm SSA}$) is expressed as

\begin{eqnarray}
\tau_{\rm SSA} =&& 1.7\times 10^4 \left(\frac{\epsilon_e}{0.1}\right) \left(\frac{\epsilon_B}{0.1}\right)^{5/4} \left(\frac{\dot{M}}{10^{-6}\,M_{\odot}\,{\rm yr^{-1}}}\right)^{43/20} \nonumber \\
&&\times \left(\frac{v_{\rm w}}{115\,{\rm km\,s^{-1}}}\right)^{-43/20} \left(\frac{E_{\rm kin}}{10^{51}\,{\rm erg}}\right)^{9/20} \nonumber \\
&&\times \left(\frac{M_{\rm ej}}{10\,M_{\odot}}\right)^{-7/20} \left(\frac{t}{10\,{\rm day}}\right)^{-18/5} \left(\frac{\nu}{1\,{\rm GHz}}\right)^{-7/2}.
\label{eq:tau_ssa}
\end{eqnarray}
If $\tau_{\rm SSA} > 1$, the optically thin synchrotron assumption of Equation (\ref{eq:ssa}) needs to be modified. In this case, the flux density of the optically thick synchrotron emission can be described as
\begin{eqnarray}
F_{\nu} =&& 4.31\times 10^{-3}\,{\rm mJy}\, \left(\frac{\dot{M}}{10^{-6}\,M_{\odot}\,{\rm yr^{-1}}}\right)^{-9/20} \nonumber \\
&&\times \left(\frac{\epsilon_B}{0.1}\right)^{-1/4} \left(\frac{v_{\rm w}}{115\,{\rm km\,s^{-1}}}\right)^{9/20} \nonumber \\
&&\times \left(\frac{E_{\rm kin}}{10^{51}\,{\rm erg}}\right)^{9/10} \left(\frac{M_{\rm ej}}{10\,M_{\odot}}\right)^{-7/10}  \nonumber \\
&&\times \left(\frac{D}{6.85\,{\rm Mpc}}\right)^{-2}\left(\frac{t}{10\,{\rm day}}\right)^{23/10} \left(\frac{\nu}{1\,{\rm GHz}}\right)^{5/2}.
\label{eq:thick}
\end{eqnarray}
The mass-loss rate corresponding to each detected flux density ($F_{\rm obs}$) was calculated using $F_{\rm obs} = F_\nu \exp{(-\tau_{\rm FFA})}$ \citep[see][]{Fransson98, Matsuoka19}. This calculation yielded two solutions. One solution is the optically thin case where $\tau_{\rm FFA} < 1$ and $\tau_{\rm SSA} < 1$, corresponding to Equation (\ref{eq:ssa}). The other is the case where $\tau_{\rm FFA}>1$ or $\tau_{\rm SSA}>1$, or both, and this case is collectively called ``optically thick'' throughout this paper.

The unfilled symbols in Figure \ref{fig:mdot} show the derived mass-loss rates corresponding to the optically thick solutions. The estimated optical depths at the JVN-detected epochs were $\tau_{\rm FFA} \simeq 3$ and $\tau_{\rm SSA} \simeq 100$, while those at the VLA-detected epoch were $\tau_{\rm FFA} \simeq 4$ and $\tau_{\rm SSA} \simeq 800$. If the JVN- and VLA- detected emissions are both optically thick, then the mass-loss rates had decreased from several $\times 10^{-3}$ to $\sim 10^{-4}\, M_{\odot}\,{\rm yr^{-1}}$, from 28 to 6 years before the explosion. 

\subsection{Mass-loss History of the Progenitor}\label{subsec:mass-loss}

Based on the optically thick and thin interpretations, we suggest that the most probable scenario is that the emission is represented by a fully optically thick emission at 29 days (at the VLA detection) and then by optically thin emissions at 152–270 days (at the JVN detections). The higher optical depth at the early time is supported by the fact that there is only non-detection data at those epochs. Furthermore, the first rising and then fading behavior observed in the light curve between 152 days and 270 days implies that the emission transitioned to optically thin.

If the emission underwent a transition from an optically thick to an optically thin state between the VLA and JVN detections, the mass-loss rate for the parameters $\epsilon_e = \epsilon_B = 0.1$ would have increased from $10^{-6}\, M_{\odot}\,{\rm yr^{-1}}$ (at the JVN-detected epochs) to $10^{-4}\, M_{\odot}\,{\rm yr^{-1}}$ (at the VLA-detected epoch) during 28 to 6 years before the explosion. The former mass-loss rate of $\sim 10^{-6}\, M_{\odot}\,{\rm yr^{-1}}$ is comparable to that of typical RSGs \citep{Goldman17}. The mass-loss rate of $\sim 10^{-4}\, M_{\odot}\,{\rm yr^{-1}}$ at the VLA-detected epoch does not agree with some upper and lower limits at $t_{\rm exp} \lesssim19\,{\rm days}$ obtained in Section \ref{subsec:mdot_constraints}. We expect that the system should be further optically thick in the earlier time, and hence, we can reject the parameter space of the mass-loss rate of $\dot{M}\lesssim 10^{-6}\,M_\odot\,{\rm yr}^{-1}$ drawn in Figure \ref{fig:mdot}. This supports the existence of a confined CSM, corresponding to the rapid enhancement of the mass-loss rate in the final few years to $\sim 10^{-2}\, M_{\odot}\,{\rm yr^{-1}}$; in this case, the synchrotron emission is fully masked by the thick confined CSM even at mm wavelengths \citep{Matsuoka19, Berger23}. 

The mass-loss rate derived from the VLA detection with the optically thick assumption is slightly inconsistent with the constraints from the KVN upper limits at $\sim 30$ days. Although the discrepancy between the observed and modeled flux densities is within a factor of a few, it indicates that our model requires a slight modification. This may entail adjusting the parameters.

Our modeling depends on some uncertain parameters. 
The value of the ejecta mass ($M_{\rm ej}$) used of $9.4\, M_{\odot}$ was derived through the modeling of the bolometric light curve by \citet{Bersten24} who estimated the zero-age main-sequence mass ($M_{\rm ZAMS}$) of $12\, M_{\odot}$. However, a wide range of $M_{\rm ZAMS}$ has been reported for SN 2023ixf, ranging from 8 to 20 $M_{\odot}$ \citep[e.g., ][]{Niu23, Soraisam23ApJ, Pledger23, Neustadt24}. According to the systematic survey of Type II SNe \citep{Martinez22}, the $M_{\rm ej}$ range corresponding to the range $8\,M_{\odot}< M_{\rm ZAMS} < 20\, M_{\odot}$ would be approximately $7\,M_{\odot}< M_{\rm ej} < 12\, M_{\odot}$. This range in $M_{\rm ej}$ results in a $\lesssim 10\%$ variation of the mass-loss rate. Another uncertain parameter is the wind velocity ($v_{\rm w}$). We employed $v_{\rm w} = 115\,{\rm km\,s^{-1}}$, derived from optical spectroscopy over the first few days, which is approximately 10 times higher than the commonly used value for RSG stars of $10\, {\rm km\, s^{-1}}$. Radio observations determine only the ratio of $\dot{M} / v_{\rm w}$, so the derived value of $\dot{M}$ will scale as $v_{\rm w}$. Additionally, $v_{\rm w}$ scales the lookback time from the explosion. Therefore, if $v_{\rm w}$ goes from $115\, {\rm km\, s^{-1}}$ to $10\, {\rm km\, s^{-1}}$, $\dot{M}$ and lookback times change by approximately a factor of 10.

The postshock energy fractions of $\epsilon_e$ and $\epsilon_B$ also have large uncertainties. Adopting the values of $\epsilon_e = 6\times 10^{-3}$ and $\epsilon_B = 5\times 10^{-2}$ derived for SN 2011dh by hydrodynamic modeling \citep{Maeda12}, the mass-loss rates under the optically thin assumption become approximately an order of magnitude higher than those for $\epsilon_e = \epsilon_B = 0.1$ (see diamonds in Figure \ref{fig:mdot}). In the optically thick case, the mass-loss rates are nearly identical to those for $\epsilon_e = \epsilon_B = 0.1$, because $\tau_{\rm FFA}$ does not depend on $\epsilon_{\rm e}$ and $\epsilon_{\rm B}$ as shown in Equation (\ref{eq:ffa}). Consequently, if the JVN-detected emission was optically thick, a decrease in mass-loss rate toward the explosion during the JVN-detected epochs must be considered, regardless of the values of $\epsilon_{\rm e}$ and $\epsilon_{\rm B}$. Assuming the emission is optically thick at the VLA-detected epoch and frequencies, and optically thin at the JVN-detected ones, the mass-loss rate had been $\sim 10^{-5}\, M_{\odot}\,{\rm yr^{-1}}$ between 28 and 46 years before the explosion, and increased to $\sim 10^{-4}\, M_{\odot}\,{\rm yr^{-1}}$ at 6 years before the explosion.

\section{Conclusions} \label{sec:conclusion}
We conducted radio follow-up observations on SN 2023ixf with the JVN, VERA, and KVN in the frequency range from 6 to 129 GHz, covering 1.7 to 269.9 days after the explosion. After non-detection at the early time ($t_{\rm exp} \lesssim 100$ days) for all observations we conducted using the JVN, VERA, and KVN, we finally detected emission in the 6.9 and 8.4 GHz bands with flux densities of $\sim 5$ mJy at $t_{\rm exp} \gtrsim 150$ days using the JVN. Comparing them with 40 $\mu$Jy measured at 29 days by the VLA \citep{Matthews23}, we infer that the flux density increased by two orders of magnitude within 120 days. The flux density reached a peak at 206 days, indicating the time till the peak in the lightcurve for SN 2023ixf is longer than that for a typical Type II SN.

The mass-loss history of the progenitor was inferred from an analytical model by assuming $M_{\rm ej} = 9.4\, M_{\odot}$ and $v_{\rm w} = 115\,{\rm km\,s^{-1}}$ derived for SN 2023ixf, and the typical values of $T_e =10^4\,{\rm K}$ and $\epsilon_e = \epsilon_B = 0.1$, or $\epsilon_e = 6\times 10^{-3}$ and $\epsilon_B = 5\times 10^{-2}$ which are derived for SN 2011dh. The most plausible scenario is that the VLA detection was fully in the optically thick regime and that the JVN detections were in transition to the optically thin emission. We estimated that the mass-loss rate has increased from $\sim 10^{-5} - 10^{-6}\, M_{\odot}\,{\rm yr^{-1}}$ to $\sim 10^{-4}\, M_{\odot}\,{\rm yr^{-1}}$ between 28 and 6 years before the explosion. Based on our mass-loss rate constraints and the suggested confined CSM structure, we propose that the mass-loss rate rapidly increased from $\sim 10^{-4}\, M_{\odot}\,{\rm yr^{-1}}$ to $\gtrsim 10^{-2}\, M_{\odot}\,{\rm yr^{-1}}$ in the final few years toward the explosion.

By conducting prompt and high-cadence radio follow-ups, we have demonstrated the capabilities of small-scale VLBIs as a time-domain astronomy facility, which will also be useful for the follow-up of radio transients found in the forthcoming Square Kilometre Array era. Future observations with detailed information on the spectrum will provide accurate constraints for the emission models. Our results indicate that for nearby SNe, observations repeated for approximately $\gtrsim 10^2$ days are important for detection. Future VLBI imaging may be able to measure the expanding motion of SN 2023ixf if the SN remains bright. Based on an assumed expansion velocity of $10^4\,{\rm km\,s^{-1}}$, the expanding motion could be resolved in 5 years with a 3 mas resolution VLBI.

\begin{acknowledgments}
We are grateful to the JVN, VERA, and VERA staff members who helped operate the telescopes and process the data. In particular, A. Yamauchi provided invaluable assistance with the data analysis for VERA. We also thank the anonymous referee for helpful comments and suggestions. VERA is a facility operated by the National Astronomical Observatory of Japan. KVN is a facility operated by the Korea Astronomy and Space Science Institute. Data analysis was in part carried out on the Multi-wavelength Data Analysis System operated by the Astronomy Data Center (ADC), NAOJ. YI was supported by the Japan Society for the Promotion of Science (JSPS) KAKENHI grant JP23K13151. TM was supported by the JSPS KAKENHI grant JP20H01904, the National Science and Technology Council, Taiwan under grant No. MOST 110-2112-M-001-068-MY3, and the Academia Sinica, Taiwan under a career development award under grant No. AS-CDA-111-M04. KM was supported by the JSPS KAKENHI grant JP20H00174 and JP24H01810. KN was supported by the JSPS KAKENHI grant JP15H00784. SCY was supported by the National Research Foundation of Korea (NRF) grant (NRF-2019R1A2C2010885). 
\end{acknowledgments}

\vspace{5mm}
\facilities{JVN, VERA, KVN}

\software{astropy \citep{Astropy13,Astropy18,Astropy22}, AIPS \citep{Greisen03}, Difmap \citep{Shepherd94}}

\bibliography{reference}{}
\bibliographystyle{aasjournal}

\end{document}